\documentclass[12pt]{article}
\usepackage{amsmath,amssymb,epsfig}

\usepackage{color}
\def\unit{{\relax{\rm 1\kern-.26em I}}}

\makeatletter

\renewcommand\section{\@startsection {section}{1}{\z@}%
                                   {-3.5ex \@plus -1ex \@minus -.2ex}%
                                   {2.3ex \@plus.2ex}%
                                   {\normalfont\large\bfseries}}

\renewcommand\subsection{\@startsection{subsection}{2}{\z@}%
                                     {-3.25ex\@plus -1ex \@minus -.2ex}%
                                     {1.5ex \@plus .2ex}%
                                     {\normalfont\normalsize\bfseries}}

\newcount\hour \newcount\minute
\hour=\time \divide \hour by 60
\minute=\time
\def\now{%
\ifnum \hour<13
  \ifnum \hour=0 \advance \hour by 12 \number\hour:\else \number\hour:\fi%
     \ifnum \minute<10 0\fi%
     \number\minute%
\ A.M.%
\else \advance \hour by -12 \number\hour:%
  \ifnum \minute<10 0\fi%
  \number\minute%
  \ P.M.%
\fi%
}

\makeatother

\begin{document}

\baselineskip=18pt  
\numberwithin{equation}{section}  
\allowdisplaybreaks  



%
%


\thispagestyle{empty}

\vspace*{-2cm}
\begin{flushright}
\end{flushright}

\begin{flushright}
\end{flushright}

\begin{center}

\vspace{1.4cm}

\vspace{0.5cm}
{\bf\Large Dyonic Catalysis in the KPV Vacuum Decay}

\vspace*{0.5cm}

\vspace*{0.5cm}

{\bf
Yuichiro Nakai$^{1}$, Yutaka Ookouchi$^{2}$ and Norihiro Tanahashi$^{3}$} \\
\vspace*{0.5cm}

${ }^{1}${\it Department of Physics and Astronomy, Rutgers University, Piscataway, NJ 08854, USA}\\
${ }^{2}${\it Faculty of Arts and Science, Kyushu University, Fukuoka 819-0395, Japan  }\\
${ }^{3}${\it Institute of Mathematics for Industry, Kyushu University, Fukuoka 819-0395, Japan  }\\

\vspace*{0.5cm}

\end{center}

\vspace{1cm} \centerline{\bf Abstract} \vspace*{0.5cm}

We investigate catalysis induced by a dyonic impurity in the metastable vacuum studied by Kachru, Pearson and Verlinde, 
which can be relevant to vacuum decay in the KKLT scenario. The impurity is a D3-brane wrapping on $ \mathbb{S}^3$ in the Klebanov-Strassler geometry. The effect of the D3-brane can be encoded in the world-volume theory of an NS5-brane as an electromagnetic field on it. As the field strength becomes large, instability of the vacuum enhances. As a result, the lifetime of the metastable vacuum becomes drastically shorter.

\newpage
\setcounter{page}{1} 



\section{Introduction}

Recent progress in string theories has been revealing that there exist a large number of metastable vacua. This involved vacuum structure of string theories is called string landscape \cite{Stringland}. Among such metastable vacua, if there is a vacuum corresponding to our universe, it has to have a small positive cosmological constant. In the celebrated work \cite{KKLT},
Kachru, Kallosh, Linde and Trivedi (KKLT)
proposed a scenario realizing de Sitter vacua in string theories.\footnote{Stability of the KKLT vacuum is still controversial. See \cite{deSitter0,deSitter1} and references therein, and also see e.g.\ \cite{recent} for recent discussions. Especially, according to the recent swampland conjecture shown in \cite{OoguriVafa1}, a de Sitter vacuum is forbidden in the string theories. In this paper, since we take the Planck mass to infinity and study the non-compact limit of the internal space, the swampland criterion can be trivially satisfied.} 
Since the KKLT vacuum is metastable, it decays within a finite time-length. There are two kinds of instabilities to the KKLT vacuum. One is destabilization of the volume-moduli and the other is annihilation of anti-D3-branes with background fluxes: To uplift the anti-de Sitter vacuum, KKLT added anti-D3-branes at the tip of the deformed conifold \cite{KKLT}, and such anti-branes can decay with the background fluxes \cite{KPV}.
In this paper, we 
discuss a catalytic effect on this latter decay process due to an impurity in the KKLT setup. Since the decay process occurs quite near the tip of the conifold, we can treat the total geometry as the non-compact Klebanov-Strassler (KS) geometry \cite{KSsolution} without losing control. This allows us to neglect gravitational effects in the four-dimensional spacetime and thus drastically simplify the analysis of the vacuum lifetime. In this non-compact limit, the system is essentially the same as the model studied by Kachru, Pearson and Verlinde (KPV) \cite{KPV}. See ref.~\cite{Frey} for an early work of a decay process of the KPV vacuum. Our goal is to examine the consequence of the D3-brane impurity and its catalytic effect in the KKLT scenario, for which we need to take into account nontrivial electromagnetic fields on the brane in the setup of \cite{KPV} as we explain below.

In our setup, the D3-brane impurity is introduced to the KKLT scenario as follows.
Near the tip of the deformed conifold, there is a non-vanishing $ \mathbb{S}^3$, and the anti-D3-branes puff up and form an NS5-brane by the Myers effect \cite{Myers}. Wrapping a D3-brane on the $\mathbb{S}^3$, 
one can introduce a point-like object which can be seen as a dyonic particle from the NS5-brane point of view. Since the 3-form RR-flux is threading $ \mathbb{S}^3$, charges of fundamental strings are induced on the wrapped D3-brane \cite{Witten}. The fundamental strings emanating from the D3-brane can end on the NS5-brane. Because of the charges induced by these strings, the object looks a dyon in four-dimensional spacetime spanned by the NS5-brane. This dyon is a soliton that has a purely stringy origin and has nothing to do with a symmetry breaking.
This kind of metastable soliton was firstly discussed in \cite{Verlinde} and later studied in various setups of string theories \cite{StringOokouchi}. In this paper, we investigate further on stringy metastable solitons and show that their existence makes the lifetime of the metastable vacuum drastically shorter. The impurity enhances the bubble nucleation rate and causes a spatially inhomogeneous decay of the vacuum. 


The idea of catalysis induced by solitons was firstly pointed out in ref.~\cite{Original} and applied to phenomenological model building later \cite{Pheno1,Pheno2}. Also, it was discussed in the context of string theory in ref.~\cite{StringOokouchi}. In this paper, we would like to go a step further to a more involved but quite interesting setup in string theory such as the KKLT model. As long as we focus on the tip of the deformed conifold, the analysis of \cite{KPV} works as is even in the KKLT model.
A difference from \cite{KPV} is that we need to take into account the nontrivial electric and magnetic fields induced on the NS5-brane due to the D3-brane impurity. 
We will study how these fields affect the tunneling rate of the metastable vacuum by employing the thin-wall approximation and numeric analysis.

The plan of the paper is following. In section 2, we briefly review the Klebanov-Strassler geometry
\cite{KSsolution}, especially near the tip of the deformed conifold, and the KPV metastable vacuum by introducing anti-D3-branes. In section 3, we numerically show dyonic solutions in the KPV metastable vacuum, which corresponds to the field configuration before the bubble nucleation.
In section 4, we study the catalytic effect induced by such dyonic objects. By using the thin-wall approximation of the solutions, we show that the lifetime of the vacuum becomes drastically shorter.
Section 5 is devoted to discussions and conclusions.

\section{Review of the Klebanov-Strassler geometry}

In this section, we briefly review the Klebanov-Strassler geometry. The authors of \cite{KSsolution} studied Type IIB string theory compactified on the deformed conifold which is a gravity dual description of the $SU(N)\times SU(N+M)$ gauge theory. The ranks $M$ and $N$ correspond to the numbers of fractional D3- and D3-brane charges which are described by the fluxes on the deformed conifold
\begin{equation}
M={1\over 4\pi \alpha^{\prime} }\int_{ \mathbb{S}^3} F_3~,\qquad N={1\over (4\pi \alpha^{\prime})^2 }\int_{ \mathbb{S}^2\times \mathbb{S}^3} F_5~.
\end{equation}
where ${\alpha^\prime}$ is the square of the string length. Since we 
focus on
the $ \mathbb{S}^3$ near the origin of the radial direction $r$ of the deformed conifold \cite{KSsolution}, let us review the metric around the origin. The radial direction has the minimum value defined by $r_{\rm min}^3\propto \epsilon^2$ at which the conifold rounds off. $\epsilon$ is the deformation parameter. As in \cite{KSsolution,Ouyang}, it 
is 
useful to introduce another parameterization $\tau$ defined by
\begin{equation}
r^2={3\over 2^{5/3}} \epsilon^{4/3} e^{2\tau/3}~.
\end{equation}
At $\tau=0$, there exists a non-vanishing $ \mathbb{S}^3$ whose metric is given by 
\begin{equation}
d\Omega_{\mathbb{S}^3}^2 = \epsilon^{4/3}(2/3)^{1/3} d\Omega_{3}^2~,
\end{equation}
where $d\Omega_{3}^2$ is the round metric of the three-dimensional sphere with unit radius. On the other hand, the remaining sub-manifold $ \mathbb{S}^2$ vanishes in the limit $\tau \to 0$ as
$d\Omega_{\mathbb{S}^2}^2 \propto \tau^2$. Thus, the metric for the non-vanishing sub-manifold at $\tau=0$ becomes \cite{KSsolution,Ouyang}
\begin{eqnarray}
ds^{ 2}_{4+3}={\epsilon^{4/3}\over  2^{1/3} c_0^{1/2} g_s M \alpha^{\prime}} dx_\mu dx^\mu+ {2\over 6^{1/3}}g_s M\alpha^{\prime} c_0^{1/2} d\Omega_{3}^2~,
\end{eqnarray}
where $c_0\simeq 0.7180$ and $g_s$ is the string coupling constant.
$\mu = 0,1,2,3$ denotes the Minkowski spacetime.
The metric can be represented as 
\begin{eqnarray}
ds^2_{4+3}=a_0^2 dx_\mu dx^\mu+ b_0^2 g_s M\alpha^{\prime}   \left(d \Psi^2 +\sin^2 \Psi d {\Omega}_{2}^2 \right) ,
\end{eqnarray}
where $d\Omega_{2}^2$ is the round metric of the two-dimensional sphere with unit radius.
We have also defined the dimensionless quantities, 
\begin{equation}
a_0^2\equiv {\epsilon^{4/3}\over  2^{1/3} c_0^{1/2} g_s M \alpha^{\prime}} \simeq 0.9366 {\epsilon^{4/3}\over  g_s M \alpha^{\prime}} \ , \qquad b_0^2\equiv {2\over 6^{1/3}} \, c_0^{1/2}\simeq 0.9327~.
\end{equation}

To facilitate the numerical analysis in this work,
let us introduce a dimensionless coordinate of the Minkowski spacetime,
\begin{equation}
\tilde{x}^{\mu}\equiv  {a_0 \over \sqrt{b_0^2g_s M{\alpha^\prime}}} \, x^{\mu}~.
\label{dimless}
\end{equation}
With this new coordinate, the metric is simply presented as 
\begin{equation}
ds_{4+3}^2=b_0^2 g_s M \alpha^{\prime} \left[  \eta_{\mu \nu }d\tilde{x}^{\mu} d \tilde{x}^{\nu}+d\Psi^2+\sin^2 \Psi d \Omega_2^2   \right].
\label{dimensionlessmetric}
\end{equation}

\section{A dyonic solution in the KPV metastable vacuum}

As was discussed in the paper \cite{KPV}, the anti-D3-branes 
added to the KS geometry
can puff-up by the Myers effect \cite{Myers} and make an NS5-brane wrapping on $ \mathbb{S}^2$ inside the non-vanishing $ \mathbb{S}^3$ at $\tau=0$. 
In the KS background, $C_0$ field is zero. Also, $B_2$ and $C_4$ fields go to zero in the limit $\tau \to 0$. Thus, according to the paper \cite{NS-five}, the total action of the NS5-brane is given by\footnote{In a small $\tau$ region, $B_2\propto \tau$ and $F_5\propto \tau$, so both two fields vanish at the origin.
It is worth noting that the Chern-Simon term $F_2 \wedge F_2 \wedge C_2$ does not contribute in the present background because this term is proportional to $C_0$ field which is vanishing. On the other hand, the term $F_2 \wedge F_2 \wedge B_2$ is allowed. However, in the limit $\tau \to 0$, the field $B_2$ goes to zero and does not contribute either. }
\begin{eqnarray}
S=-{T_{NS}\over g_s^2}\int d^6\xi \sqrt{-{\rm det} \left(g_{ab}+2\pi g_s \alpha^{\prime}\widetilde{{\cal F}} \right)} - T_{NS} \int B_6~,
\label{NS5action0}
\end{eqnarray}
where $T_{ NS}$ is the tension of the NS5-brane and $2\pi \alpha^{\prime} \widetilde{\cal F}=2\pi \alpha^{\prime} F_2 -C_2$.

Now, we introduce an impurity by wrapping a D3-brane on $\mathbb{S}^3$ at $\tau=0$. Since the RR 3-form flux threads $\mathbb{S}^3$, a charge of the fundamental string is induced on the wrapped D3-brane \cite{Witten}. To reconcile the charge conservation for the induced charge, we have to introduce the fundamental string ending on the D3-brane and the NS5-brane. This object can be seen as a dyonic particle from the viewpoint of Minkowski spacetime spanned by the NS5-brane.
When the metastable vacuum decays in this setup, the D3-brane forms a bound state with the domain wall created by the decay as follows. 
At the domain wall, the NS5-brane sweeps a portion of $\mathbb{S}^3$ between the loci corresponding to the metastable vacuum and the true vacuum. Thus the domain wall NS5-brane and the dyonic D3-brane are on the top of each other on $\mathbb{S}^3$.
In this case, the D3-brane dissolves into the NS5-brane to form a bound state \cite{Text},
and the effect of the D3-brane 
manifests in the Lagrangian as the electromagnetic field on the NS5-brane. 

Now we are ready to consider the Lagrangian describing the NS5-brane. Let us discuss the electromagnetic field in the dimensionless coordinate
\begin{equation}
F_{\mu \nu}dx^{\mu} \wedge dx^{\nu}=\left({ \alpha^{\prime}{b_0^2g_s M} \over a_0^2}\right)  F_{\mu \nu}d\tilde{x}^{\mu} \wedge d\tilde{x}^{\nu}=\widetilde{F}_{\mu \nu}d\tilde{x}^{\mu} \wedge d\tilde{x}^{\nu}~.
\end{equation}
The diagonal block corresponding to Minkowski spacetime of the matrix $\tilde{g}_{ab}+2\pi g_s \alpha^{\prime} \tilde{F}_{ab}$ can be represented as 
\begin{eqnarray}
b_0^2g_s M \alpha^{\prime}\times 
\begin{pmatrix} 
  (-1+\dot{\Psi}^2) &  \dot{\Psi}\Psi^{\prime}+\left({2\pi g_s \alpha^{\prime} \over a_0^2}\right) E  &0 & 0  \\ 
  \dot{\Psi}\Psi^{\prime}-\left({2\pi g_s  \alpha^{\prime} \over a_0^2}\right) E & (1+\Psi^{\prime 2}) &0 & 0 \\ 
  0 & 0 & \tilde{r}^2 & \left({2\pi g_s \alpha^{\prime}  \over a_0^2}\right) B\sin \theta   \\ 
 0 & 0 &-\left({2\pi g_s \alpha^{\prime} \over a_0^2}\right) B\sin \theta  & \tilde{r}^2\sin^2 \theta \\ 
 \end{pmatrix} \nonumber ~,
\end{eqnarray}
 where 
$\tilde{r}$ is the radial coordinate in the dimensionless coordinates (\ref{dimless}), and 
the NS5-brane is wrapping $\mathbb{S}^2$ at $\Psi = \Psi(\tilde t,\tilde r)$.
The dot and prime denote $\tilde t$ and $\tilde r$ derivatives, respectively. 
$B$ stands for the magnetic field multiplied by $\tilde{r}^2$, which is a constant and proportional to the number of the D3-branes added as the impurity. $E$ is the electric field induced on the brane. 

The diagonal block of the matrix $\tilde{g}_{ab}+2\pi g_s \alpha^{\prime} \tilde{F}_{ab}$ corresponding to the internal space $d\Omega_2^2$ in (\ref{dimensionlessmetric}) spanned by $\theta_I,\  \phi_I$ is given by
\begin{eqnarray} 
\begin{pmatrix} 
 b_0^2g_s M \alpha^{\prime} \sin^2 \Psi & \alpha^{\prime} \Big(\pi g_s p -g_s M (\Psi-{1\over 2}\sin 2 \Psi)\Big)\sin \theta_I  \\ 
 - \alpha^{\prime} \Big(\pi g_s p -g_s M (\Psi-{1\over 2}\sin 2 \Psi)\Big)\sin \theta_I  & b_0^2g_s M \alpha^{\prime} \sin^2\Psi \sin^2 \theta_I  \\ 
 \end{pmatrix} \nonumber ~,
\end{eqnarray}
where $p$ is the number of the anti-D3-branes.
Thus, the DBI action for the NS5-brane becomes
\begin{eqnarray}
S_{\rm DBI }&=&-{T_{NS}\over g_s^2} \int d^6 \xi \sqrt{-{\rm det} (g_{ab}+2\pi g_s \alpha^{\prime}\widetilde{\cal F})} \nonumber \\
&=&-{\mu_5 (b_0^2g_sM \alpha^{\prime})^2 \over g_s^2\alpha^{\prime\, 3} } (4\pi)^2 g_sM\alpha^{\prime} \int d\tilde{t} d\tilde{r} \sqrt{1-\dot{\Psi}^2+{\Psi}^{\prime 2} - {\cal E}^2}  \sqrt{\tilde{r}^4+{\cal B}^2} \times \nonumber \\
&& \sqrt{b_0^4\sin^4 \Psi  +\left({\pi p\over M}- \left(\Psi-{1\over 2} \sin 2\Psi\right)\right)^2 } ~,
\end{eqnarray}
where we defined $\mu_5=T_{NS} \alpha^{\prime\, 3}$. In the second line, the integrals over angular coordinates $(\theta_I ,\ \phi_I)$ and $(\theta, \, \phi)$ are performed. Here, we defined the dimensionless fields by
\begin{equation}
{\cal E}\equiv {2\pi g_s \alpha^{\prime} \over a_0^2} E \ , \qquad {\cal B}\equiv {2\pi g_s \alpha^{\prime} \over a_0^2} B~.
\end{equation}
Finally we define the dimensionless action  ${S}_{\rm DBI }= -16\pi^3g_s M^3b_0^4 \mu_5  \widetilde{S}_{\rm DBI }$,
\begin{eqnarray}
\widetilde{S}_{\rm DBI }= \int d\tilde{t} d\tilde{r} \sqrt{1-\dot{\Psi}^2+{\Psi}^{\prime 2} - {\cal E}^2} \sqrt{\tilde{r}^4+{\cal B}^2} \, {1\over \pi} \sqrt{b_0^4\sin^4 \Psi  +\left({\pi p\over M}- \left(\Psi-{1\over 2} \sin 2\Psi\right)\right)^2 }~. \nonumber 
\end{eqnarray}

\begin{figure}[htbp]
\centering
\includegraphics[width=.47\linewidth]{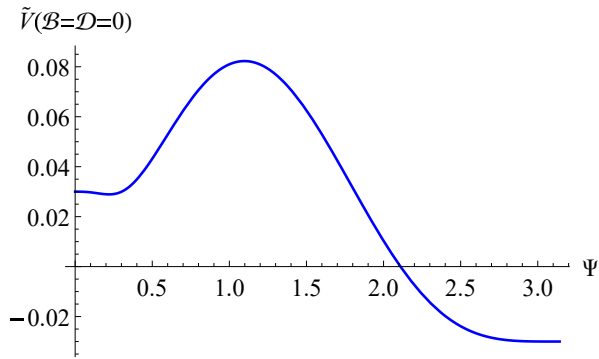}
\caption{ \sl \small
Potential energy for a static NS5-brane with ${\cal B} = {\cal D}= 0$. Besides the global potential minimum is at $\Psi = \pi$, there is a local minimum at $\Psi = 0.220$, which is approximately given by $\Psi = 2 \pi p / b_0^4 M$ when $p/M \ll 1$ \cite{KPV}.}
\label{fig:potential0}
\end{figure}

Next, we move on to the Chern-Simons term in (\ref{NS5action0}). By using the KS solution \cite{KSsolution}, it can be written as 
\begin{equation}
S_{\rm CS}=-T_{NS}\int B_6={\mu_5\over g_s  \alpha^{\prime\, 3}} \int (dV)_4 \int_{S^2} C_2 ~.
\end{equation}
In the dimensionless coordinate, the volume form is represented as
\begin{equation}
({dV})_4=a_0^4  d^4x =a_0^4 \left({ \sqrt{b_0^2g_sM{\alpha^\prime}} \over a_0} \right)^4\tilde{r}^2 \sin \theta \, d \theta \, d \phi \, d\tilde{r} \, d\tilde{t}~.
\end{equation}
Plugging back into the action, we obtain
\begin{eqnarray}
S_{\rm CS}=-T_{NS}\int B_6&=&{\mu_5 \over g_s \alpha^{\prime\, 3} } 4\pi a_0^4\left({ \sqrt{b_0^2g_sM{\alpha^\prime}} \over a_0} \right)^4\alpha^{\prime}  \int d\tilde{t} d\tilde{r} \tilde{r}^2 (4\pi M)\left(\Psi-{1\over 2}\sin 2\Psi \right) \nonumber \\
&=&
16\pi^2 g_s M^3  b_0^4 \mu_5
\int d\tilde{r} d\tilde{t} \, \tilde{r}^2\left(\Psi-{1\over 2}\sin 2\Psi \right)~.\nonumber 
\end{eqnarray}
In total, the action is%
\footnote{
Using Eq.~(\ref{dimless}), it can be checked that Eq.~(\ref{V2}) with ${\cal B}={\cal E}=0$ coincides with Eq.~(4.9) of \cite{KPV} up to a factor $1/\alpha'^2$, which is set to the unity therein.}
\begin{eqnarray}
S&=&S_{\rm DBI}+S_{\rm CS} \nonumber \\
&=&\int d\tilde{t} d\tilde{r}  \left[ -16\pi^3g_s M^3b_0^4 \mu_5  \sqrt{1-\dot{\Psi}^2+{\Psi}^{\prime 2} - {\cal E}^2}  \sqrt{\tilde{r}^4+{\cal B}^2} \, V_2(\Psi) \right. \nonumber \\
&&\left. \qquad\quad~
+
16\pi^2 g_s M^3  b_0^4  \mu_5
\tilde{r}^2\left(\Psi -{1\over 2}\sin 2\Psi\right) \right]~,
\end{eqnarray}
where we defined $V_2(\Psi)$ by
\begin{equation}
V_2(\Psi)={1\over \pi} \sqrt{b_0^4\sin^4 \Psi  +\left({\pi p\over M}- \left(\Psi-{1\over 2} \sin 2\Psi\right)\right)^2 }~.
\label{V2}
\end{equation}
With this function, the total action can be written as
\begin{eqnarray}
{S}&=&16\pi^3g_s M^3b_0^4 \mu_5 \widetilde{S} 
=
16\pi^3g_s M^3b_0^4 
\mu_5 \int d\tilde{t} d\tilde{r} \widetilde{\cal L} ~.
\label{NS5action}
\end{eqnarray}
Here, we defined the dimensionless Lagrangian
\begin{eqnarray}
\widetilde{{\cal L}}&=& -V_2 (\Psi)\sqrt{1-\dot{\Psi}^2+{\Psi}^{\prime 2} - {\cal E}^2}  \sqrt{\tilde{r}^4+{\cal B}^2} + {\tilde{r}^2  \over \pi}\left(\Psi -{1\over 2}\sin 2\Psi\right)  
\nonumber 
~.
\end{eqnarray}
$\cal E$ depends on $\tilde r$, hence it is convenient to change the variable that is independent of $\tilde r$. This can be accomplished by the Legendre transformation in terms of the electric displacement ${\cal D} \equiv \frac{\partial \tilde{\cal L}}{\partial {\cal E}}$ \cite{Emparan,KlebanovPufu}. ${\cal D}$ is proportional to the number of the 3-form fluxes $M$ and that of the D3-branes added as the impurity. Explicit form of the electric displacement is given by
\begin{eqnarray}
{\cal D}\equiv  {\partial \widetilde{\cal L}\over \partial {\cal E}} =V_2(\Psi) {\cal E} \sqrt{\tilde{r}^4+{\cal B}^2 \over 1-\dot{\Psi}^2+\Psi^{\prime\,  2}-{\cal E}^2}~.
\end{eqnarray}
By solving the equation, the electric field can be written as  
\begin{equation}
{\cal E}=\sqrt{{\cal D}^2(1-\dot{\Psi}^2+\Psi^{\prime\, 2})\over V_2(\Psi)^2(\tilde{r}^4+{\cal B}^2)+{\cal D}^2}~.
\end{equation}
Then the new Lagrangian which is a function of ${\cal B}$ and ${\cal D}$ is given by
\begin{equation}
\widetilde{{\cal L}}_D=\widetilde{\cal L}-{\cal D}{\cal E}=-\sqrt{V_2(\Psi)^2(\tilde{r}^4+{\cal B}^2)+{\cal D}^2}\sqrt{1-\dot{\Psi}^2+\Psi^{\prime\,  2}}+{\tilde{r}^2\over \pi}  \left(\Psi -{1\over 2}\sin 2\Psi\right)  ~.
\label{tildeLagD}
\end{equation}
As a consistency check, let us consider the action with ${\cal B}={\cal D}=0$. In this case, we can rewrite the action as
\begin{equation}
\widetilde{S}={1\over 4\pi}\int d \tilde{r} d \tilde{t} d \theta d \phi  \, \tilde{r}^2 \sin \theta \left[-V_2(\Psi)\sqrt{1-\dot{\Psi}^2+\Psi^{\prime\,  2}}+{1\over \pi}  \left(\Psi -{1\over 2}\sin 2\Psi\right)  \right] ~.
\end{equation}
Once we put $\Psi'=0$, this action coincides with that for a homogeneous configuration shown in \cite{KPV}. From this action, 
the potential energy for a static configuration is given by
\begin{equation}
\widetilde{V}({\cal B}={\cal D}=0) \propto
V_2(\Psi) - {1\over \pi}  \left(\Psi -{1\over 2}\sin 2\Psi\right) .
\label{potential0}
\end{equation}
As shown in figure~\ref{fig:potential0},
this potential has a true vacuum at $\Psi = \pi$ and
a metastable point at $\Psi \sim 2 \pi p / b_0^4 M$ \cite{KPV}. Also, by taking $\dot{\Psi}={\cal D}=0$ and ${\cal B} \neq 0$, we can reproduce the action shown in \cite{Verlinde}.  

In this work we study the tunneling from this metastable vacuum toward the true vacuum.
Once $\cal B$ or $\cal D$ is turned on, the potential energy becomes $\tilde r$ dependent but still it has a structure similar to (\ref{potential0}).
The metastable vacuum of this $\tilde r$ dependent potential 
corresponds to the field configuration before the tunneling studied in section~\ref{sec:thinwall}.

\begin{figure}[htbp]
\begin{center}
 \includegraphics[width=.47\linewidth]{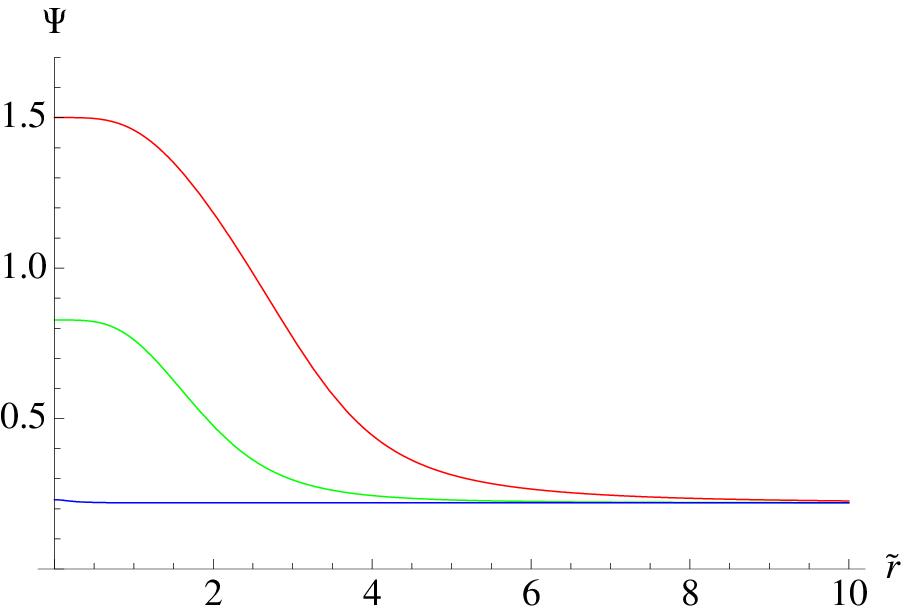} \hspace{0.7cm}
  \includegraphics[width=.47\linewidth]{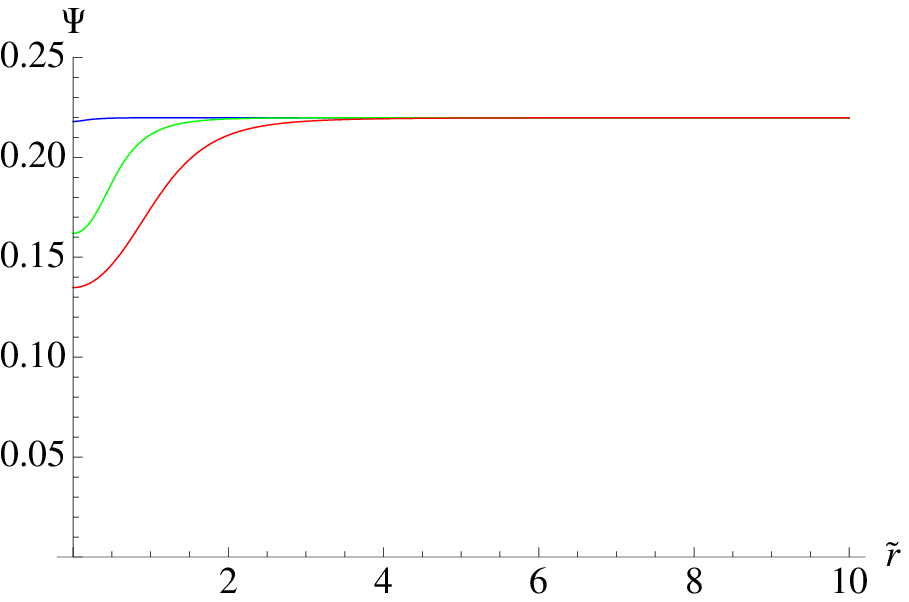} 
\vspace{-.1cm}
\caption{\sl \small Plots of static solutions
for $p/M=0.03$. 
%
The left panel shows the purely electric solutions with ${\cal B}=0$ where the red, green and blue lines correspond to ${\cal D}=1$, $0.3$ and $0.001$.
The right panel shows purely magnetic solutions with ${\cal D}=0$, where the red, green and blue lines correspond to ${\cal B}=1.4$, $0.3$ and $0.001$.
}
\label{DyonicSolution}
\end{center}
\end{figure}

Now let us study 
inhomogeneous ($\Psi'\neq 0$)
static solutions 
imposing the regularity at the center
$\Psi'(\tilde r = 0)= 0$ following \cite{KlebanovPufu}.
These solutions correspond to the metastable configuration before the tunneling process occurs.
In figure \ref{DyonicSolution}, we show numerical solutions of such static configurations
obtained using the relaxation method.
At large $\tilde{r}$, contributions of ${\cal D}$ and ${\cal B}$ become effectively negligible, and then the NS5-brane 
resides at
the KPV metastable vacuum. 
As the radius becomes small, 
the NS5-brane position $\Psi$ becomes closer to $\Psi = \pi$ for a purely 
electric
solutions 
while it becomes closer to $\Psi = 0$ for a purely 
magnetic
solutions.
This behavior can be understood as follows.
Expanding the Lagrangian of the brane to the leading order in ${\cal D}$ and ${\cal B}$, we find
\begin{equation}
\widetilde{S} 
\supset 2\pi \int d^4 \tilde{x} \, \left[ V_2(\Psi) \left( { {\cal B} \over 4\pi \tilde{r}^2}\right)^2+   {1\over V_2(\Psi)}      \left( {{\cal D}^2 \over 4\pi \tilde{r}^2} \right)^2  \right]~,
\label{tildeH}
\end{equation}
where we recovered the angular integrals. Let us focus on the purely magnetic case first, for which only the first term is present in the above.
We see that the energy density due to nonzero $\cal B$ becomes larger as $\tilde r$ becomes smaller.
To compensate this energy increase, $\Psi$ is forced to shift to smaller value since
$V_2(\Psi)$, defined by (\ref{V2}), is an increasing function of $\Psi$. On the other hand, in the purely electric case, $V_2(\Psi)$ appears as the denominator of the second term, hence $\Psi$ becomes larger as $\tilde r$ becomes smaller to make the total energy smaller.

From the four-dimensional point of view, this phenomenon is caused because
the electric permittivity and 
the magnetic permeability depend on the value of $\Psi$ in a particular way.
Comparing (\ref{tildeH}) with the energy density of the classical electromagnetism
\begin{equation}
{1\over 2 \varepsilon} \, { \mathbb{D}\cdot \mathbb{D} }+{1\over 2 \mu} \,  { \mathbb{B}\cdot \mathbb{B} }~,
\end{equation}
we find the relations $\varepsilon \propto V_2(\Psi)$ and $\mu \propto V_2(\Psi)^{-1}$.

It is worth noting that in the present setup, the 3-form flux $M$ is a large number because the curvature of the conifold should be large enough to make the supergravity approximation reliable. When we wrap the D3 brane on the $ \mathbb{S}^3$, $M$ units of the fundamental charge are induced on it, which means that ${\cal D}$ is proportional to $M$. Therefore, in our assumption, ${\cal D}$ is larger than ${\cal B}$. In this case, from figure \ref{DyonicSolution}, we expect that profile functions of the dyonic particles should be much similar to the ones in the left panel of the figure. For this type of dyonic solutions,
$\Psi$ is shifted to larger value around the center.
This feature would enhance
the phase transition from the metastable vacuum 
since the configuration of the NS5-brane is pushed toward that of the true vacuum $\Psi = \pi$.
In the next section, we will confirm that the tunneling rate to the true vacuum is indeed enhanced when nonzero $\cal B$ and $\cal D$ are present.

\section{Numerical study of the decay rate}
\label{sec:thinwall}

In the previous section, we discussed that the NS5-branes are bent near the origin of $\Psi$ when a dyonic particle exists at the origin. In this section, by assuming the existence of a stable solution for an appropriate choice of the parameters ${\cal B}$ and ${\cal D}$, we estimate the decay rate by 
applying the thin-wall approximation to the bubble.

\subsection{Thin-wall approximation}

The goal of this section is to construct solutions describing the phase transition from the metastable KPV vacuum to the true vacuum.
An obstacle for it is that the Lagrangian (\ref{tildeLagD}) depends on both $\tilde t$ and $\tilde r$, hence one would need to solve two-dimensional partial differential equations to obtain solutions corresponding to the 
phase transition.
To simplify this problem, we employ the thin-shell approximation for the domain wall, with which the problem is reduced to solving an ordinary differential equation.
In the thin-wall limit, the profile of a domain wall solution is given by
\begin{equation}
\Psi
=
(\Psi_{\rm max}-\Psi_{\rm min}) 
\left[
1-\theta\bigl(
\tilde{r}-R(\tilde{t})
\bigr) 
\right]
+\Psi_{\rm min}~.
\end{equation}
where $\theta$ is the step function\footnote{To be precise, the value $\Psi_{\rm max}$ is not constant. The profile function before tunneling should be something like functions shown in figure \ref{DyonicSolution}. Clearly, we see $\Psi_{\rm max}<\pi$ for the initial profile. On the other hand, 
well after the tunneling $\Psi$ should converge to $\Psi= \pi$, which corresponds to the true vacuum. Hence
we have to treat  $\Psi_{\rm max}$ as a time-dependent function ideally. However, for the sake of simplicity, we assume  $\Psi_{\rm max}=\pi$. When $\cal D$ is large and $\Psi_{\rm max}$ is close to $\pi$ in the initial profile function, our calculation becomes reliable. }. 
The NS5-brane annihilates with the background flux at the bubble wall, hence the electromagnetic field on the NS5-brane should be zero inside the bubble. Namely, we set ${\cal B}={\cal D}=0$ for $\tilde{r}<R$. 
For this ansatz, the differentials of $\Psi(\tilde t, \tilde r)$ are written by the delta function,
\begin{equation}
{\partial \Psi \over \partial \tilde{r}}=-(\Psi_{\rm max}-\Psi_{\rm min}) \delta \bigl(\tilde{r}-R(\tilde{t})\bigr)\ , \quad {\partial \Psi \over \partial \tilde{t}}=(\Psi_{\rm max}-\Psi_{\rm min})\dot{R}(\tilde{t}) \delta \bigl(\tilde{r}-R(\tilde{t})\bigr)
~,
\end{equation}
and then the kinetic part of the Lagrangian (\ref{tildeLagD}) is approximated as
\begin{equation}
\sqrt{1-\dot{\Psi}^2+\Psi^{\prime\, 2}}\simeq (\Psi_{\rm max}-\Psi_{\rm min}) \sqrt{1-\dot{R}^2} \, \delta \bigl(\tilde{r}-R(\tilde{t})\bigr)~.
\end{equation}
For this ansatz, it is useful to divide the radial direction $\tilde{r}$ into three intervals $[0,R]$, $[R, R+\Delta \tilde{r}]$ and $[R, R_{\infty}]$, each of which corresponds to the bubble interior, the bubble wall region and the bubble exterior.
Applying the above approximation to (\ref{tildeLagD}), the action for each interval is given by 
\begin{eqnarray}
 \widetilde{S}^{\bf 1st}&=&\int d\tilde{t} \left[ -V_2(\Psi_{\rm max})\int_0^R d \tilde{r} \, \tilde{r}^2+{R^3 \over 3\pi  }\left( \Psi_{\rm max}-{1\over 2}\sin 2\Psi_{\rm max} \right) \right] ~,  \nonumber \\
\widetilde{S}^{\bf 2nd}&=&\int d\tilde{t} \left[ \int_{\Psi_{\rm max}}^{\Psi_{\rm min}} d \Psi \sqrt{V_2(\Psi)^2 (R^4+{\cal B}^2)+{\cal D}^2 }\sqrt{1-\dot{R}^2}\right]~, \\
\widetilde{S}^{\bf 3rd}&=&\int d\tilde{t} \left[ -V_2(\Psi_{\rm min})\int_R^{R_{\infty}} d \tilde{r} \sqrt{\tilde{r}^4+{\cal B}^2+{{\cal D}^2 \over V_2^2(\Psi_{\rm min})}  }+{R^3_{\infty}-R^3 \over 3\pi  }\left( \Psi_{\rm min}-{1\over 2}\sin 2\Psi_{\rm min} \right) \right] ~. \nonumber 
\end{eqnarray}
To make the total 
action finite, it is convenient to subtract the action for the static solution $R(\tilde{\tau})=0$,
\begin{eqnarray}
\widetilde{S}^{\infty}&=&\int d\tilde{t} \left[ -V_2(\Psi_{\rm min})\int_0^{R_{\infty}} d \tilde{r} \sqrt{\tilde{r}^4+{\cal B}^2+{{\cal D}^2 \over V_2^2(\Psi_{\rm min})}  }+{R^3_{\infty} \over 3\pi  }\left( \Psi_{\rm min}-{1\over 2}\sin 2\Psi_{\rm min} \right) \right. \nonumber \\
&&\left. 
+\int_{\Psi_{\rm max}}^{\Psi_{\rm min}} d \Psi \,  \sqrt{V_2(\Psi)^2 {\cal B}^2+{\cal D}^2   }\right]~.
\end{eqnarray}
The total action becomes
\begin{eqnarray}
{\widetilde{S}^{\rm tot}}&\equiv & \widetilde{S}^{\bf 1st}+\widetilde{S}^{\bf 2nd}+\widetilde{S}^{\bf 3rd}-\widetilde{S}^{\infty}\nonumber \\
&=& \int d\tilde{t} \left[ -V_2(\Psi_{\rm max})  \int_0^R d \tilde{r} \, \tilde{r}^2+{R^3 \over 3\pi  }\left( \Psi_{\rm max}-{1\over 2}\sin 2\Psi_{\rm max} \right) \right. \nonumber \\
&& 
+V_2(\Psi_{\rm min})\int_0^R d \tilde{r} \sqrt{\tilde{r}^4+{\cal B}^2+{{\cal D}^2 \over V_2^2(\Psi_{\rm min})}  }-{R^3 \over 3\pi  }\left( \Psi_{\rm min}-{1\over 2}\sin 2\Psi_{\rm min} \right) \nonumber \\
&& 
+ \left. \int_{\Psi_{\rm max}}^{\Psi_{\rm min}} d \Psi \left(\sqrt{V_2(\Psi)^2 (R^4+{\cal B}^2)+{\cal D}^2 }\sqrt{1-\dot{R}^2} -\sqrt{V_2(\Psi)^2 {\cal B}^2+{\cal D}^2 } \right) \right] ~.
\end{eqnarray}
To see the physical meaning, let us consider the potential energy of a static configuration
\begin{align}
4\pi \widetilde{V}^{\rm tot} &=
- \Delta V {4\pi R^3 \over 3}-4\pi V_2 (\Psi_{\rm min})\int_0^R d \tilde{r} \left( \sqrt{\tilde{r}^4+{\cal B}^2+{{\cal D}^2 \over V_2^2(\Psi_{\rm min})}  } -\tilde{r}^2 \right) \nonumber \\
&\quad
 +  4\pi \int^{\Psi_{\rm max}}_{\Psi_{\rm min}} d \Psi \left(\sqrt{V_2(\Psi)^2 (R^4+{\cal B}^2)+{\cal D}^2 }-\sqrt{V_2(\Psi)^2 {\cal B}^2+{\cal D}^2 } \right)  ~,
\end{align}
where we defined $\Delta V ={\cal V}(\Psi_{\rm min}) -{\cal V}(\Psi_{\rm max})$ with ${\cal V}(\Psi)={1\over 4\pi}[V_2(\Psi) - {1\over \pi}  \left(\Psi -{1\over 2}\sin 2\Psi\right)]$. The first term is the 
energy deficit due to
the true vacuum inside the bubble. 
The second term is the 
energy deficit due to
disappearance of the electromagnetic fields inside the bubble. 
The third term corresponds to the surface energy of the bubble originating from the tension and electromagnetic fields on it. 

\subsection{Bounce action}

\begin{figure}[htbp]
\begin{center}
 \includegraphics[width=.45\linewidth]{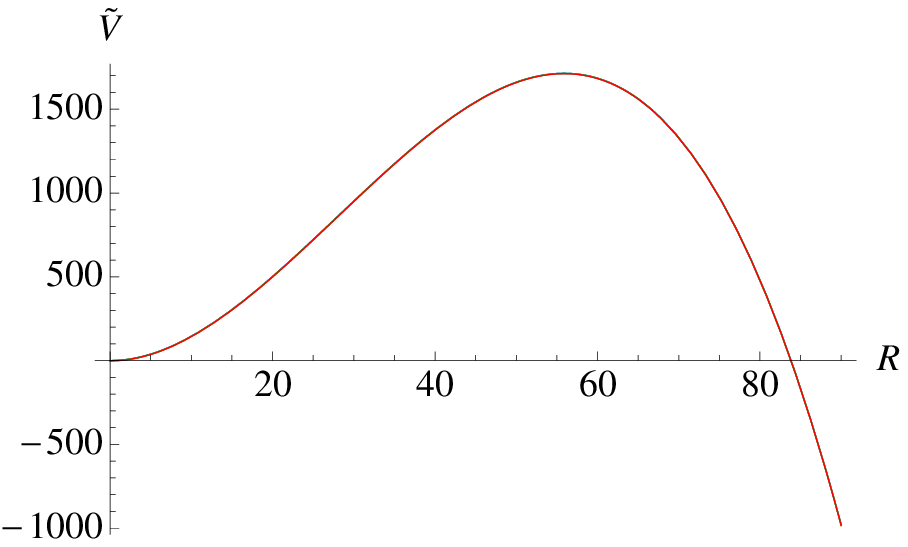} \hspace{0.7cm}
  \includegraphics[width=.45\linewidth]{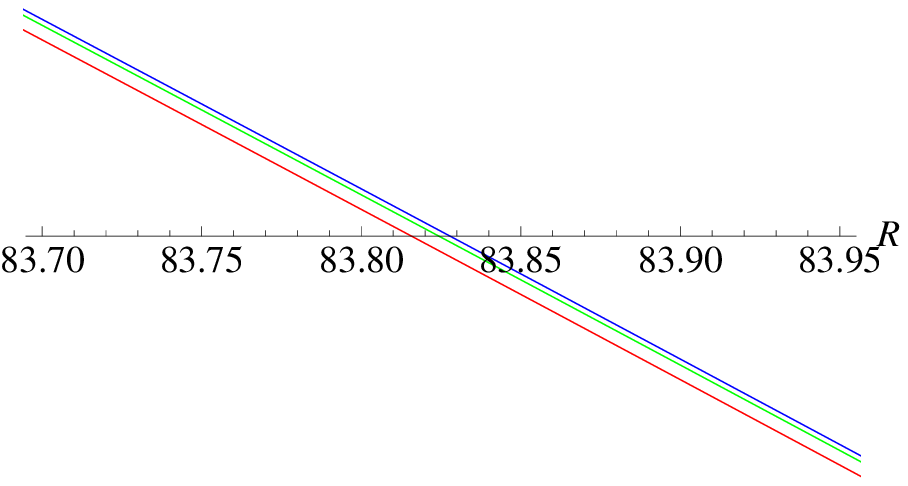}
\vspace{-.1cm}
\caption{\sl \small Plots of the dimensionless potential $\widetilde{V}$ defined in \eqref{effpotential} with 
$p/M=0.03$. 
We choose ${\cal D}=0$ and the blue, green and red lines correspond to ${\cal B}=0.01$, $0.3$ and $1$ respectively. In the right panel, we magnify the functions around $R=83.8$. 
}
\label{PotentialA}
\end{center}
\end{figure}

\begin{figure}[htbp]
\begin{center}
 \includegraphics[width=.45\linewidth]{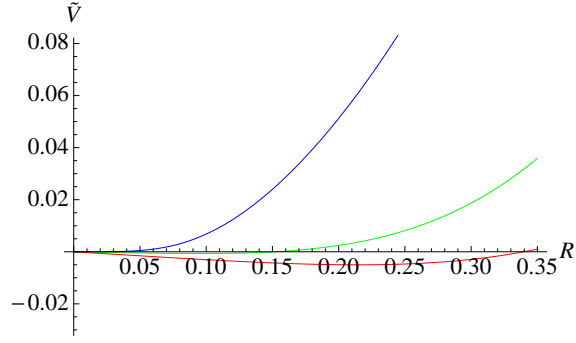} 
\vspace{-.1cm}
\caption{\sl \small Plots of the dimensionless potential $\widetilde{V}$ defined in \eqref{effpotential} with 
$p/M=0.03$. 
We choose ${\cal D}=0$ and the blue, green and red lines correspond to ${\cal B}=0.01$, $0.3$ and $1$ respectively. We magnify the functions around $R=0.15$. 
Each curve has a local minimum at $R = R_{\rm ini} > 0$, at which the domain wall before the vacuum decay resides.}
\label{PotentialB}
\end{center}
\end{figure}

\begin{figure}[bt]
\begin{center}
 \includegraphics[width=.45\linewidth]{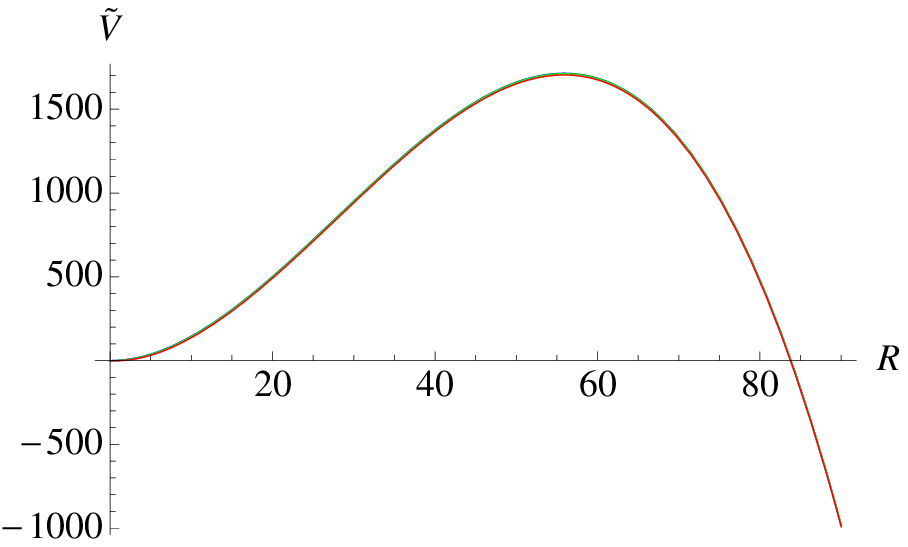} \hspace{0.7cm}
  \includegraphics[width=.45\linewidth]{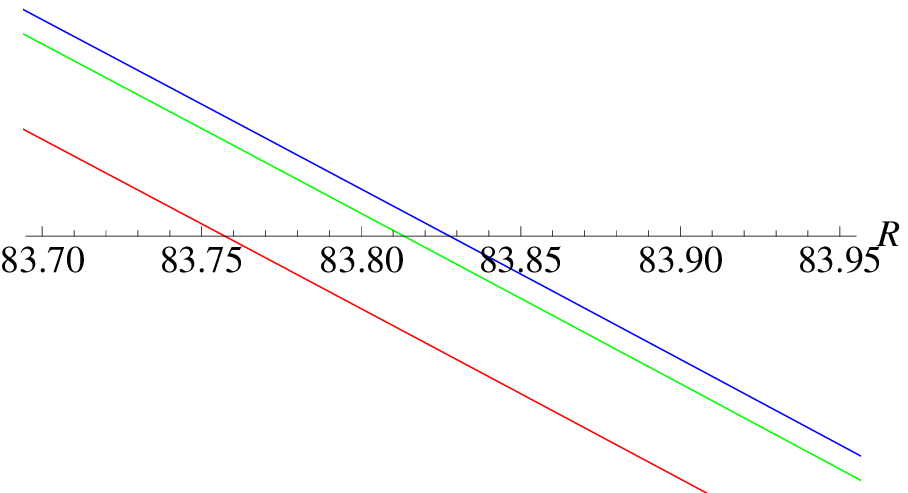}
\vspace{-.1cm}
\caption{\sl \small Plots of the dimensionless potential $\widetilde{V}$ defined in \eqref{effpotential} with 
$p/M=0.03$. 
We choose ${\cal B}=0$ and the blue, green and red lines correspond to ${\cal D}=0.01$, $0.3$ and $1$ respectively. In the right panel, we magnify the functions around $R=83.8$. }
\label{PotentialC}
\end{center}
\end{figure}

Now, we are ready to study a catalytic decay of the KPV metastable vacuum. To estimate the decay rate, we use Coleman's method \cite{Coleman} and proceed basically along the lines of \cite{Emparan}.  Let us first introduce functions defined by 
\begin{eqnarray}
&&T(R,{\cal B}, {\cal D}) \equiv -\int_{\Psi_{\rm max}}^{\Psi_{\rm min}} d\Psi \sqrt{V_2(\Psi)^2(R^4+{\cal B}^2) +{\cal D}^2}~,  \nonumber \\
&& H(R,{\cal B}, {\cal D})\equiv {R^3 \over 3\pi} \left(\Psi_{\rm max}-\Psi_{\rm min} -{1\over 2} \sin 2\Psi_{\rm max}+{1\over 2} \sin 2\Psi_{\rm min} \right)+T(R=0, {\cal B}, {\cal D}) \nonumber \\
&&
\hphantom{H(R,{\cal B}, {\cal D})\equiv}
-V_2(\Psi_{\rm max}) \int_0^R d \tilde{r} \ \tilde{r}^2+V_2(\Psi_{\rm min}) \int_0^R d \tilde{r} \sqrt{\tilde{r}^4+ {\cal B}^2 +{ {\cal D}^2 \over V_2(\Psi_{\rm min})^2 }}
\label{TH}
~.
\end{eqnarray}
With these functions, the Euclidean action can be written as
\begin{equation}
\widetilde{S}_E^{\rm tot}= \int d\tilde{\tau} \left[  -H(R,{\cal B}, {\cal D}) +T(R,{\cal B}, {\cal D})\sqrt{1+\dot{R}^2}  \right]~.
\label{Stot}
\end{equation}
When $p/M$ is small, which is necessary for neglecting the back-reaction of the anti-D3-branes,
the angular coordinate at the metastable vacuum is approximately given by 
$\Psi_{\rm min}={2\pi p / b_0^4M}$ \cite{KPV}. 
As for the maximum value of $\Psi$, we simply assume the value for the supersymmetric vacuum, namely $\Psi_{\rm max}=\pi$. 
Below, we show the dimensionless effective potential defined by
\begin{equation}
\widetilde{V}=-H(R,{\cal B},{\cal D})+ T(R,{\cal B},{\cal D})
\label{effpotential}
\end{equation}
for several values of ${\cal B}$ and ${\cal D}$. In figures \ref{PotentialA} and \ref{PotentialB}, we choose ${\cal D}=0$, 
$p/M=0.03$. 
The blue, green and red lines correspond to ${\cal B}=0.01$, $0.3$ and $1$ respectively. From the figure \ref{PotentialB}, one sees that there exists a metastable point at nonzero $R$. We denote this minimum $R_{\rm ini}$, 
hence $\partial_R \widetilde V(R_{\rm ini})=0$. 
We consider a tunneling process from $R=R_{\rm ini}$ to $R=R_*$, where $R_*$ is defined by $\widetilde V(R_*) = \widetilde V(R_{\rm ini})$.
Figure~\ref{PotentialC} shows the plots for ${\cal B}=0$ and 
$p/M=0.03$. 
The blue, green and red lines correspond to ${\cal D}=0.01$, $0.3$ and $1$. 

By using the initial condition given by $R = R_{\rm ini}$ and $\dot{R}=0$ at $\tilde{\tau}=0$, the conserved Hamiltonian can be represented as 
\begin{equation}
{T(R, {\cal B}, {\cal D}) \over \sqrt{1+\dot{R}^2}} =H(R, {\cal B}, {\cal D})-H(R_{\rm ini}, {\cal B}, {\cal D})+T(R_{\rm ini}, {\cal B}, {\cal D})~.
\end{equation}
Solving in $\dot{R}$, we obtain
\begin{equation}
\dot{R}=\sqrt{T(R, {\cal B}, {\cal D})^2-(H(R, {\cal B}, {\cal D})+K_0   )^2 \over (H(R, {\cal B}, {\cal D})+K_0 )^2 }~,
\end{equation}
where we defined 
\begin{equation}
K_0=-H(R_{\rm ini}, {\cal B}, {\cal D})+T(R_{\rm ini}, {\cal B}, {\cal D}) ~.
\end{equation}
Also, to obtain the bounce action we subtract the action for the static solution $R=R_{\rm ini}$
\begin{eqnarray}
{1\over 2}\widetilde{B}_b &=&\widetilde{S}_E-\widetilde{S}_E (R_{\rm ini}) \nonumber \\
&=& \int d \tilde{\tau} \left[ -H(R,{\cal B}, {\cal D}) +T(R, {\cal B}, {\cal D}) \sqrt{1+\dot{R}^2 }-K_0 \right] \nonumber \\
&=& \int_{R_{\rm ini}}^{R_{*}} d R \sqrt{T(R, {\cal B}, {\cal D})^2-(H(R,{\cal B}, {\cal D})+K_0)^2 }  ~.
\label{B}
\end{eqnarray}

\begin{figure}[htbp]
\begin{center}
 \includegraphics[width=.47\linewidth]{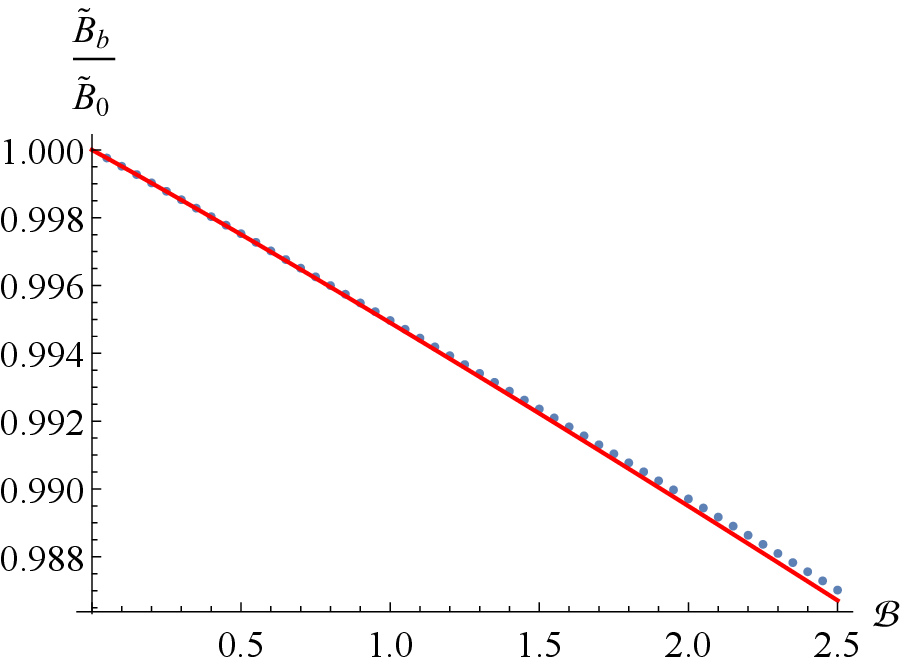}\hspace{0.7cm}
  \includegraphics[width=.47\linewidth]{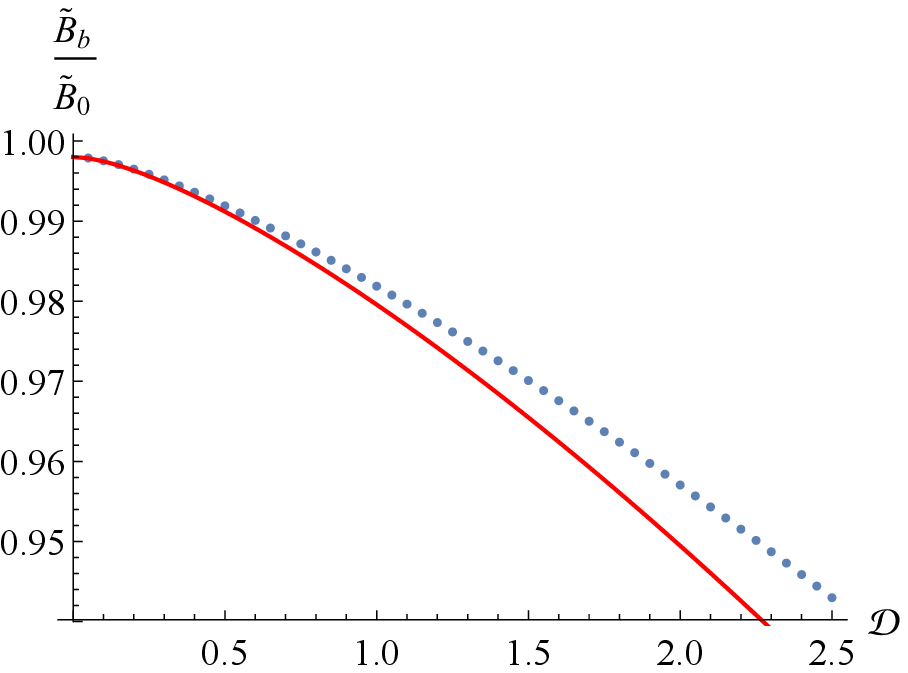}
\vspace{-.1cm}
\caption{\sl \small Numerical values of the bounce action for $p/M=0.08$ 
normalized by $\widetilde{B}_0$, which is the bounce action for  ${\cal B}={\cal D}=0$.
The blue dots show the numerical results, and the red curves show the analytic results based on Eq.~(\ref{I(delta)}) that is valid for ${\cal B}, {\cal D}\ll 1$. Equation~(\ref{I(delta)}) reduces to Eq.~(\ref{Bfinal}) in the limit ${\cal B}, {\cal D}, p/M \to 0$.
 In the left panel, we show the $\cal B$ dependence of the bounce action for ${\cal D}=0$. In the right panel, we show the $\cal D$ dependence when ${\cal B}=0.4$. In both cases, as ${\cal B}$ or ${\cal D}$ increases, the bounce action becomes smaller. 
}
\label{figFinA}
\end{center}
\end{figure}

In figure \ref{figFinA}, we show numerical values of the bounce action (\ref{B}) 
and its dependence on ${\cal B}$ and ${\cal D}$ for $p/M=0.08$.
The left panel shows the $\cal B$ dependence of the bounce action for ${\cal D}=0$, and the right panel shows the $\cal D$ dependence when ${\cal B}=0.4$. 
In both cases, the bounce action decreases monotonically as ${\cal B}$ or ${\cal D}$ increases.
%
The bounce action decreases linearly with respect to ${\cal B}$ when ${\cal D}=0$ (left panel), while it shows more complicated behavior when both $\cal B$ and $\cal D$ are turned on (right panel).
%
%
%
To understand this behavior it is useful to expand the bounce action for small ${\cal B}, {\cal D}$, which results in (see Appendix~\ref{appA} for details)
\begin{equation}
\frac12 \widetilde B_b({\cal B},{\cal D})
=
\frac{27\pi}{128}
\frac{\left(
\int_{\Psi_{\rm min}}^{\Psi_{\rm max}} 
V_2(\Psi)
d\Psi 
\right)^4}{(p/M)^3}
\left(
1 - 
\frac{64}{9\pi} \left(\frac{p}{M}\right)^2
\frac{\int_{\Psi_{\rm min}}^{\Psi_{\rm max}} 
V_2(\Psi) 
\sqrt{{\cal B}^2 + \frac{{\cal D}^2}{V_2^2(\Psi)}}
d\Psi}{\left(
\int_{\Psi_{\rm min}}^{\Psi_{\rm max}} 
V_2(\Psi)
d\Psi
\right)^3}
+ \cdots
\right)~.
\label{Bfinal}
\end{equation}
This expression implies that the bounce action decreases linearly with respect to $\sqrt{{\cal B}^2 + \frac{{\cal D}^2}{V_2^2(\Psi)}}$ when ${\cal B}, {\cal D}$ are small, and this is the origin of the ${\cal B}, {\cal D}$ dependence of the bounce action mentioned above.
This expression also implies that the bounce action is more sensitive to $\cal D$ rather than $\cal B$, because the former is multiplied by a factor $1/V_2(\Psi)$, which varies from ${\cal O}(1)$ (when $\Psi\sim \pi$) to ${\cal O}(M/p)\gg 1$ (when $\Psi\sim 0$).
%

The tunneling probability is an exponential of the bounce action, hence we may conclude that nonzero $\cal B$ and $\cal D$ enhances the tunneling probability significantly. The $\cal B$ and $\cal D$ fields are nothing but the manifestation of the D3-brane impurities we introduced, and in this sense 
one concludes that the KPV vacuum is efficiently catalyzed by such impurities to decay to the true vacua.

\section{Discussions and conclusions}

In this paper, 
we focused on
the decay of metastable vacua in Type IIB string theory and investigated the catalytic effect induced by D3-branes wrapped on $ \mathbb{S}^3$ at the tip of the deformed conifold. 
We first studied the bound state of the D3-brane and domain wall NS5-brane which connects the metastable vacuum to true vacuum. We found that a dyonic particle induces instability of the metastable state near the particle. Then, we estimated the decay rate employing the Coleman's method and the thin-wall approximation. We showed that the life-time of the metastable vacuum becomes shorter when non-vanishing electromagnetic field is present. We also pointed out that this type of the vacuum decay may occur even in the de Sitter vacuum in the KKLT scenario if D3-brane impurities are present.
Here we stress that our result is independent from the KKLT model and applies to any other models associated with conifolds with flux. 

Once the decay occurs, the cosmological constant becomes negative and hence such a process must be suppressed within our observed universe. 
For an expanding universe with Hubble constant $H$ with the decay rate $\Gamma$, the decay probability within the Hubble time is roughly estimated as $H^{-4}\Gamma$ and it must be much smaller than the unity~\cite{Turner:1992tz}. The decay rate is roughly estimated as $\Gamma \sim l_s^{-4} \exp(-B_b)$, where $l_s$ is the string length scale. When ${\cal B}={\cal D}=0$, $\Gamma$ is given by
\begin{equation}
 \Gamma \sim 
l_s^{-4} \exp\left[
-\frac{27\pi^4}{4} g_s M^3 b_0^4 \mu_5 \frac{\left(\int_{\Psi_\text{min}}^{\Psi_\text{max}} V_2(\Psi)\right)^4}{(p/M)^3}
\right]\,.
\end{equation}
Then the upper bound on the decay probability $H^{-4}\Gamma \equiv \epsilon \ll 1$ is translated to a constraint on $p/M$ as
\begin{equation}
 \frac{p}{M} = 
\frac{1}{\left(\log\epsilon^{-1}\right)^{1/3}}
\left(
\frac{
\frac{27\pi^4}{4}g_s M^3 b_0^4 \mu_5 \left(\int_{\Psi_\text{min}}^{\Psi_\text{max}} V_2(\Psi)\right)^4
}{
\log\left(
H^{-4} l_s^{-4}
\right)
}
\right)^{1/3}.
\end{equation}
When ${\cal B}, {\cal D}$ are nonzero, $p/M$ must be decreased further
since $\widetilde B_b$ is a function decreasing with respect to ${\cal B}, {\cal D}$ as we showed in section~\ref{sec:thinwall}.
To make this constraint more precise, we would need to evaluate the prefactor of $\Gamma$ (see e.g.~\cite{Garriga:1993fh}) specifying the cosmological scenario and the value of $H$ \cite{Garriga:1993fh}. It would be interesting to pursue this issue based on some stringy inflation models.

In this work we neglected the gravitational effects in the four-dimensional spacetime by taking the decoupling limit and focusing on the tip of the deformed conifold. It is desirable to improve our analysis taking the gravitational effect into account, so that we can study the influences of the vacuum decay discussed in this work to the de Sitter universe realized in the KKLT scenario. 
For example,
catalysis of the phase transition due to the black holes and compact objects were discussed in ref.~\cite{Gregory:2013hja} taking the gravitational effect into account.
Though the ``catalyst'' in our setup is a stringy particle and qualitatively different from theirs, it would be fruitful to make connection between these catalytic processes to gain deeper insight into the phenomenology in the early universe and to find observational evidence of the background theory governing it.

%

This decay process is associated with a dyonically charged spherical domain wall, and it might be interesting to examine 
its observational signature in
our universe. In some cases, such spherical domain walls collapse to form black holes.
Studying the dynamics and observational consequences of such spherical domain walls employing techniques of, e.g., ref.~\cite{Maeda:1985ye} would be one of possible future directions of our study.

\section*{Acknowledgement}

We are grateful to Minoru Eto for useful discussions. YN is supported by the DOE grant DE-SC0010008. YO would like to thank Rutgers University for their hospitality. YO and NT are supported by Grant-in-Aid for Scientific Research from the Ministry of Education, Culture, Sports, Science and Technology, Japan (No.17K05419, No.18H01214 and No.18K03623)
and AY 2018 Qdai-jump Research Program of Kyushu University.

\appendix 

\section{Analytic estimate of the $\cal B$, $\cal D$ dependence of the bounce action}
\label{appA}

We derive an analytic expression of the bounce action when $\cal B$, $\cal D$ are small.

\subsection{Definitions}

As mentioned in section~\ref{sec:thinwall}, 
We assume $p/M \ll 1$ and then $\Psi_{\rm min}, \Psi_{\rm max}$ are approximated as
\begin{equation}
 \Psi_{\rm min}=\frac{2\pi p}{b_0^4M}~,
\qquad
 \Psi_{\rm max}= \pi~.
\end{equation}
Then the bounce action $\widetilde{B}_b$ is given by Eq.~(\ref{B}).
$T(R,{\cal B}, {\cal D})$ and $H(R,{\cal B},{\cal D})$ 
defined by Eq.~(\ref{TH})
may be expressed as
\begin{align}
T(R,{\cal B}, {\cal D}) 
&
=
\int_{\Psi_{\rm min}}^{\Psi_{\rm max}} 
V_2(\Psi)\sqrt{R^4+\rho^4(\Psi)} d\Psi 
~,  
\\
H(R,{\cal B}, {\cal D})
&=
\frac{R^3}{3}
\left(
\frac{\Delta\Psi}{\pi} - V_2(\Psi_{\rm max})
\right)
+ V_2(\Psi_{\rm min})\int_0^R \sqrt{\tilde r^4 + \rho^4(\Psi_{\rm min})}d\tilde r 
+ \int_{\Psi_{\rm min}}^{\Psi_{\rm max}}  V_2(\Psi) \rho^2(\Psi) d\Psi
~.
\end{align}
In the above expressions, we introduced $\Delta \Psi$ and $\rho(\Psi)$ defined by
\begin{align}
 \Delta \Psi &\equiv 
\Psi_{\rm max}-\Psi_{\rm min} -{1\over 2} \sin 2\Psi_{\rm max}+{1\over 2} \sin 2\Psi_{\rm min}
=
\pi + {\cal O}\left(\left(\frac{p}{M}\right)^3\right)~,
\\
\rho^4(\Psi) &\equiv
{\cal B}^2 + {{\cal D}^2 \over  V_2(\Psi)^2 }~.
\end{align}
$V_2(\Psi)$ at $\Psi = \Psi_{\rm min}, \Psi_{\rm max}$ are estimated as
\begin{equation}
V_2(\Psi_{\rm min}) = \frac{p}{M} + {\cal O}\left(\left(\frac{p}{M}\right)^3\right)~,
\qquad
V_2(\Psi_{\rm max}) = 1 - \frac{p}{M}~.
\end{equation}

\subsection{${\cal B} = {\cal D}=0$}

Before studying the general case, we summarize the expressions for ${\cal B} = {\cal D}=0$.
In this case, 
$\rho(\Psi)$ vanishes and then 
$T,H$ and $\widetilde{B}_b$
are given by
\begin{align}
T(R,0, 0)
&=
\left(
\int_{\Psi_{\rm min}}^{\Psi_{\rm max}} 
V_2(\Psi)
d\Psi 
\right)R^2
~,  
\\
H(R,0,0)
&=
\frac13
\left(
\frac{\Delta\Psi}{\pi} - V_2(\Psi_{\rm max})
+ V_2(\Psi_{\rm min})
\right)
R^3
\simeq
\frac{2p}{3M} R^3
~,
\label{approxH}
\\
K_0 &= 0~,
\\
\widetilde V(R) &= 
- H(R,0,0) + T(R,0,0)
=
-\frac{2p}{3M} R^3 + \left(
\int_{\Psi_{\rm min}}^{\Psi_{\rm max}} 
V_2(\Psi)
d\Psi 
\right)R^2~,
\end{align}
where the last expression of $H(R,0,0)$~(\ref{approxH}) follows in the limit $p/M\to 0$.
From $\widetilde V(R)$ we find 
\begin{equation}
 R_{\rm ini}({\cal B}=0, {\cal D}=0) = 0~,
\qquad
R_*({\cal B}=0, {\cal D}=0) = 
\frac{3}{2(p/M)}
\int_{\Psi_{\rm min}}^{\Psi_{\rm max}} 
V_2(\Psi)
d\Psi ~.
\end{equation}
Using these expressions the bounce action is calculated as
\begin{align}
 {1\over 2}\widetilde{B}_b &=
\int_{R_{\rm ini}}^{R_{*}} d R \sqrt{T(R, {\cal B}, {\cal D})^2-(H(R,{\cal B}, {\cal D})+K_0)^2 }
\notag \\
&=
\int_0^{R_*} d R \sqrt{
\left(
\int_{\Psi_{\rm min}}^{\Psi_{\rm max}} 
V_2(\Psi)
d\Psi 
\right)^2
R^4
-\left(
\frac{2p}{3M} R^3
\right)^2
}
=
\frac{27\pi}{128}
\frac{\left(
\int_{\Psi_{\rm min}}^{\Psi_{\rm max}} 
V_2(\Psi)
d\Psi 
\right)^4}{(p/M)^3}~.
\end{align}
Also, the integral of $V_2(\Psi)$ can be evaluated for small $p/M$ as, 
setting $b_0^2 \simeq 0.9327$,
\begin{equation} 
\int_{\Psi_{\rm min}}^{\Psi_{\rm max}} 
V_2(\Psi)
d\Psi = 
1.71 - 2.29 \times \frac{p}{M} +{\cal O}\left(\left(\frac{p}{M}\right)^2\right)~.
\end{equation}

\subsection{Small ${\cal B}, {\cal D}$ }

In this section, we estimate how the bounce action is modified when small $\cal B$ and $\cal D$ are turned on.
We assume $\rho(\Psi)$ and $\rho(\Psi_{\rm min})$ are of the same order at any $\Psi$.
Then, the integrand appearing in $T$ and $H$ are roughly approximated as
\begin{equation}
\sqrt{\tilde r^4 + \rho^4(\Psi)} 
\simeq
\begin{cases}
\rho^2(\Psi) & (\tilde r \ll \rho(\Psi)) \\
\tilde r^2 & (\tilde r \gg \rho(\Psi))
\end{cases}~.
\end{equation}
Using this approximation, $T$, $H$ and $K_0$ at the leading order are estimated as
\begin{align}
T(R,{\cal B}, {\cal D}) 
&=
\int_{\Psi_{\rm min}}^{\Psi_{\rm max}} 
V_2(\Psi)\sqrt{R^4+\rho^4(\Psi)} d\Psi 
\simeq
\begin{cases}
\int_{\Psi_{\rm min}}^{\Psi_{\rm max}} 
V_2(\Psi)\rho^2(\Psi)
d\Psi
& (R \ll \rho(\Psi))
\\
\left(
\int_{\Psi_{\rm min}}^{\Psi_{\rm max}} 
V_2(\Psi)
d\Psi
\right)
R^2
& (R \gg \rho(\Psi))
\end{cases}
~,  
\\
H(R,{\cal B}, {\cal D})
&=
\frac{R^3}3
\left(
\frac{\Delta\Psi}{\pi} - V_2(\Psi_{\rm max})
\right)
+ \int_{\Psi_{\rm min}}^{\Psi_{\rm max}}  V_2(\Psi) \rho^2(\Psi) d\Psi
+ V_2(\Psi_{\rm min})\int_0^R \sqrt{\tilde r^4 + \rho^4(\Psi_{\rm min})}d\tilde r~,
\\
K_0({\cal B},{\cal D}) &
=
-\left[
\frac13 \left(\frac{\Delta \Psi}{\pi} -V_2(\Psi_{\rm max})\right) R_{\rm ini}^3
+ V_2(\Psi_{\rm min}) R_{\rm ini}\rho^2(\Psi_{\rm min})
\right]~,
\end{align}
where
\begin{equation}
\int_0^R \sqrt{\tilde r^4 + \rho^4(\Psi_{\rm min})}d\tilde r
\simeq
\begin{cases}
\rho^2(\Psi_{\rm min}) R
& (R\ll \rho(\Psi_{\rm min}))
\\
\rho^3(\Psi_{\rm min})
+ \frac13 \left[\tilde r^3\right]_{\rho(\Psi_{\rm min})}^R
=
\frac13 \left(R^3 + 2\rho^3(\Psi_{\rm min})\right)
& (R\gg \rho(\Psi_{\rm min}))
\end{cases}~.
\end{equation}
In the above expression $R_{\rm ini}$ is shifted from the value for ${\cal B} = {\cal D} = 0$.
Solving $\partial_R \widetilde{V}(R_{\rm ini}) =0$  
and $\widetilde V(R_{\rm ini}) = \widetilde V(R_*)$
assuming $R_{\rm ini} \ll \rho(\Psi) \ll 1$, we find
\begin{align}
R_{\rm ini}({\cal B}, {\cal D}) &=
\left[
\int_{\Psi_{\rm min}}^{\Psi_{\rm max}} 
\frac{2 V_2(\Psi) / V_2(\Psi_{\rm min})}{\rho^2(\Psi) \rho^2 (\Psi_{\rm min})}
d\Psi 
\right]^{-1/3} = {\cal O}\left(\rho^{4/3}\right)
~,
\label{Rini_approx}
\\
R_*({\cal B}, {\cal D}) &= 
\frac{3}{2(p/M)}
\int_{\Psi_{\rm min}}^{\Psi_{\rm max}} 
V_2(\Psi)
d\Psi 
-
\frac{2p}{3M}
\frac{
\int_{\Psi_{\rm min}}^{\Psi_{\rm max}} 
V_2(\Psi) \rho^2(\Psi)
d\Psi 
}{\left(
\int_{\Psi_{\rm min}}^{\Psi_{\rm max}} 
V_2(\Psi)
d\Psi 
\right)^2}~.
\label{R*_approx}
\end{align}

Using the above expressions, 
the differences 
$\Delta T \equiv T(R,{\cal B},{\cal D}) - T(R,0,0)$ 
and
$\Delta H \equiv H(R,{\cal B},{\cal D}) - H(R,0,0)$ 
are found to be 
\begin{align}
\Delta T &\simeq
\begin{cases}
\int_{\Psi_{\rm min}}^{\Psi_{\rm max}} 
V_2(\Psi)\rho^2(\Psi)
d\Psi
-
\left(
\int_{\Psi_{\rm min}}^{\Psi_{\rm max}} 
V_2(\Psi)
d\Psi
\right)R^2
& (R \ll \rho(\Psi))
\\
{\cal O}\left(\rho^4(\Psi)\right)
& (R \gg \rho(\Psi))
\end{cases}~,
\\
\Delta H &\simeq
 \int_{\Psi_{\rm min}}^{\Psi_{\rm max}}  V_2(\Psi) \rho^2(\Psi) d\Psi
+ V_2(\Psi_{\rm min})
\times
\begin{cases}
\rho^2(\Psi_{\rm min}) R
& (R\ll \rho(\Psi_{\rm min}))
\\
C
\rho^3(\Psi_{\rm min})
& (R\gg \rho(\Psi_{\rm min}))
\end{cases}~,
\label{DeltaH}
\end{align}
where $C\equiv \frac23 (-1)^{1/4}\left[K(-1)-i K(2)\right]\simeq 1.236$ and $K(x)$ is the complete elliptic integral of the first kind.
Among these terms, 
it turns out that 
the first term 
and the $C \rho^3\left(\Psi_{\rm min}\right)$ term in (\ref{DeltaH}) give the leading and subleading contributions to the bounce action.
Hence the bounce action with small $\rho(\Psi)$ may be approximated as 
\begin{equation}
\frac12 \widetilde B_b({\cal B},{\cal D})
\simeq
\int_{R_{\rm ini}
(\delta)
}^{R_*
(\delta)
}
dR
\sqrt{
T^2(R,0,0) 
- \left(H(R,0,0) 
+\delta
\right)^2
}
=
\int_{R_{\rm ini}
(\delta)
}^{R_*
(\delta)}
dR
\sqrt{
\alpha^2
R^4
 - \left(
\beta
R^3
+ 
\delta
\right)^2
}
~,
\label{B(delta)}
\end{equation}
where
\begin{align}
\alpha &\equiv
\int_{\Psi_{\rm min}}^{\Psi_{\rm max}} 
V_2(\Psi)
d\Psi~,
\qquad
\beta\equiv
\frac{R^3}3
\left(
\frac{\Delta\Psi}{\pi} - V_2(\Psi_{\rm max})
\right)~,
\notag
\\
\delta &\equiv 
\int_{\Psi_{\rm min}}^{\Psi_{\rm max}}  V_2(\Psi) \rho^2(\Psi) d\Psi
+ C\, V_2(\Psi_{\rm min})\rho^3(\Psi_{\rm min})~,
\label{parameters}
\end{align}
and $R_{\rm ini}(\delta)$ and $R_*(\delta)$ are redefined as the radii at which the integrand of Eq.~(\ref{B(delta)}) vanishes.

Below, we evaluate the integral on the right-hand side of Eq.~(\ref{B(delta)}) for small $\delta$,
$R_{\rm ini}(\delta)$ and $R_*(\delta)$ are roots of $\sqrt{\alpha^2 R^4 - \left(\beta R^3 + \delta\right)^2}=0$, and they are given by\footnote{%
While $R_*$ obtained from formula (\ref{R_approx2}) for the parameters (\ref{parameters}) coincides with Eq.~(\ref{R*_approx}),
$R_{\rm ini}$ does not match with Eq.~(\ref{Rini_approx}) due to the definition for $R_{\rm ini}(\delta)$ used here.
Error of 
$\widetilde B_b({\cal B},{\cal D})$
due to this mismatch is higher order in $\delta$ and negligible.}
\begin{equation}
R_{\rm ini}(\delta) \sim \frac{\delta^{1/2}}{\alpha^{1/2}}~,
\qquad
R_*(\delta) \sim
\frac{\alpha}{\beta} - \frac{\beta}{\alpha^2} \delta~.
\label{R_approx2}
\end{equation}
Then, 
Eq.~(\ref{B(delta)})
is estimated as
\begin{equation}
\frac12 \widetilde B_b({\cal B},{\cal D})
\simeq
\int_{\frac{\delta^{1/2}}{\alpha^{1/2}}}^{\frac{\alpha}{\beta} - \frac{\beta}{\alpha^2} \delta}  
dR
\sqrt{\alpha^2 R^4 - \left(\beta R^3 + \delta\right)^2}
=
\frac{\pi \alpha^4}{16 \beta^3}
\left(
1 - \frac{16}{\pi} \frac{\beta^2}{\alpha^3} \delta
\right)
+{\cal O}\left(\delta^{3/2}\right)\,.
\label{I(delta)}
\end{equation}
%
%
The approximate expression for the bounce action $\widetilde B_b({\cal B},{\cal D})$ can be obtained by plugging these parameters into Eq.~(\ref{I(delta)}).
In the limit $p/M\to 0$ and $\rho\to 0$, Eq.~(\ref{I(delta)}) reduces to Eq.~(\ref{Bfinal}).

%
%

\end{document}